\documentclass[12pt]{article}
\usepackage{geometry}                % See geometry.pdf to learn the layout options. There are lots.
\usepackage{xcolor,colortbl}

\usepackage[displaymath, mathlines]{lineno}
\newcommand*\patchAmsMathEnvironmentForLineno[1]{%
  \expandafter\let\csname old#1\expandafter\endcsname\csname #1\endcsname
  \expandafter\let\csname oldend#1\expandafter\endcsname\csname end#1\endcsname
  \renewenvironment{#1}%
     {\linenomath\csname old#1\endcsname}%
     {\csname oldend#1\endcsname\endlinenomath}}% 
\newcommand*\patchBothAmsMathEnvironmentsForLineno[1]{%
  \patchAmsMathEnvironmentForLineno{#1}%
  \patchAmsMathEnvironmentForLineno{#1*}}%
\AtBeginDocument{%
\patchBothAmsMathEnvironmentsForLineno{equation}%
\patchBothAmsMathEnvironmentsForLineno{align}%
\patchBothAmsMathEnvironmentsForLineno{flalign}%
\patchBothAmsMathEnvironmentsForLineno{alignat}%
\patchBothAmsMathEnvironmentsForLineno{gather}%
\patchBothAmsMathEnvironmentsForLineno{multline}%
}

\usepackage{graphicx}
\usepackage{amsmath}
\usepackage{amssymb}
\usepackage{epstopdf}
\usepackage{amsthm}
\usepackage{float}
\usepackage{color}
\usepackage{todonotes}
\usepackage{cite}
\usepackage[section]{placeins}
\usepackage{bbm}
 \usepackage{setspace}

\definecolor{LightCyan}{rgb}{0.88,1,1}
\newcommand{\ob}[1]{\left(#1\right)} %% produces pair of open brackets 
\newcommand{\cb}[1]{\left[#1\right]} %% produces pair of closed brackets

\newcommand{\R}{\mathbb{R}}

\newcommand{\ip}[1]{\left<#1\right>}

\newcommand{\set}[1]{\{#1\}}

\begin{document}
%\maketitle

\begin{titlepage}
  \vspace*{0cm}
  \begin{flushleft}
    \Large\bfseries
     Bone Remodeling as a Spatial Evolutionary Game  \par
  \end{flushleft}
  \vspace{0cm}
  \begin{flushleft}
   Marc D. Ryser$^{1}$ and Kevin A. Murgas$^2$
  \end{flushleft}
  \vspace{0cm}
   \begin{flushleft}\small
   \begin{enumerate}
    \item  Department of Mathematics, Duke University, Durham, NC, USA
\item Department of Biomedical Engineering, Duke University, Durham, NC, USA
\end{enumerate}
\vspace{2cm}
%{\bf Running title:} Dynamics of field cancerization\\
{\bf Corresponding author:} {\it Marc D. Ryser}, Department of Mathematics, Duke University, 120 Science Drive, 117 Physics Building, Durham, NC 27708. email: ryser@math.duke.edu, phone: +1-919-660-2847, fax: +1-919-660-2821.\\
\vspace{1cm}
%{\bf Financial support:} This work was supported by the National Institutes of Health (R01-GM096190); the Swiss National Science Foundation (P300P-154583); and the National Science Foundation (DMS-0943760).\\
\vspace{1cm}
{\bf Potential conflicts of interest:} The authors disclose no potential conflicts of interest.\\

\vspace{2cm}
%{\bf Word count abstract:} XX\\
%{\bf Word count main text:} XX\\
%{\bf Total number of tables and figures:} x (x figures, x table)\\
  \end{flushleft}
  \begin{center}
    \date{\today}
  \end{center}
\end{titlepage}

  \newpage 
  
  \section*{Abstract}

Bone remodeling is a complex process involving cell-cell interactions, biochemical signaling and mechanical stimuli. Early models of the biological aspects of remodeling were non-spatial and focused on the local dynamics at a fixed location in the bone. Several spatial extensions of these models have been proposed, but they generally suffer from two limitations: first, they are not amenable to analysis and are computationally expensive, and second, they neglect the role played by bone-embedded osteocytes. To address these issues, we developed a novel model of spatial remodeling based on the principles of evolutionary game theory. The analytically tractable framework describes the spatial interactions between zones of bone resorption, bone formation and quiescent bone, and explicitly accounts for regulation of remodeling by bone-embedded, mechanotransducing osteocytes. Using tools from the theory of interacting particle systems we systematically classified the different dynamic regimes of the spatial model and identified regions of parameter space that allow for coexistence of resorption, formation and quiescence, as observed in physiological remodeling. In coexistence scenarios, three-dimensional simulations revealed the emergence of sponge-like bone clusters. Comparison between spatial and non-spatial dynamics revealed substantial differences and suggested a stabilizing role of space. Our findings emphasize the importance of accounting for spatial structure and bone-embedded osteocytes when modeling the process of bone remodeling. Thanks to the lattice-based framework, the proposed model can easily be coupled to a mechanical model of bone loading.\\

{\bf Keywords:} Bone physiology; trabecular remodeling; osteocytes; osteoclasts, osteoblasts; spatial evolutionary games; interacting particle systems.

\newpage
%%%%%%%%%%%%%%%%%%%%%%%%%
\section{Introduction}
%%%%%%%%%%%%%%%%%%%%%%%%%

Bone remodeling is a complex mechano-biological process that is critical for maintenance of the healthy skeleton \cite{robling2006biomechanical}. During bone remodeling, bone-resorbing osteoclasts remove old and damaged bone while bone-matrix producing osteoblasts generate new bone tissue to restore structural integrity, see Figure 1A. The recruitment of osteoclasts, and subsequently osteoblasts, is mediated by bone-embedded, mechano-sensing osteocytes, which translate load-induced mechanical strains into signals to control the adaptive  remodeling process \cite{raggatt2010cellular}. Disruption of the interactions between the key cellular components of remodeling can lead to pathological states. Such is the case in osteoporosis, where hormonal changes during menopause cause imbalances in the remodeling process and can lead to fracture-prone bones, and in Paget's disease,  a condition where bone undergoes cycles of uncontrolled resorption and formation \cite{feng2011disorders}. \\
\begin{figure}[htb!]
    \centering
 \includegraphics[width=1\textwidth]{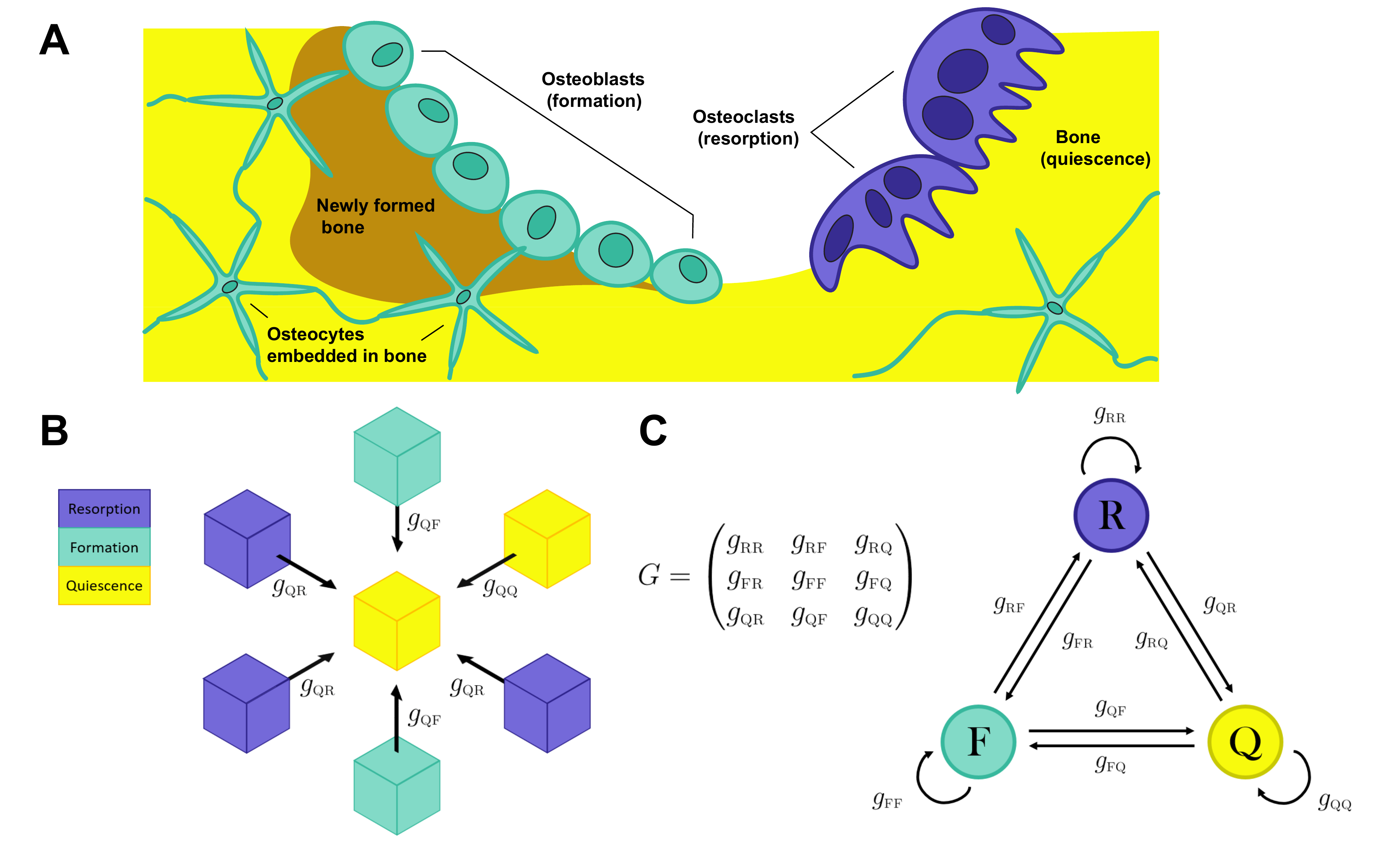} 
   \vspace{-.4cm}
   \caption{ \footnotesize  {\bf Bone Remodeling as a Spatial Evolutionary Game.} {\bf (A)} Bone remodeling is a complex multicellular process necessary for maintenance and adaptation of a healthy skeleton. Bone resorbing osteoclasts (purple) remove old and damaged bone (yellow), and osteoblasts (green, round) produce new bone matrix. Once osteoblasts have completed their task of producing new bone, they either die or become embedded in the bone tissue where they differentiate into osteocytes (green, star-shaped). Osteocytes are connected through a complex network and are thought to play an integral role in sensing bio-mechanical stimuli and translating them into chemical signals to orchestrate the remodeling process by osteoclasts and osteoblasts. {\bf (B)}. In the spatial setting, the expansion rate of a zone (center) is determined by the constitution of its neighbors and the corresponding interaction strengths $g_{XY}$. Note that $g_{XY}$ quantifies the impact of a zone of type $Y$ on a zone of type $X$. {\bf (C)}  The interactions between resorption (R), formation (F) and quiescence (Q). This network determines the inter-species interactions in the evolutionary game theory model. The $3\times 3$ pay-off matrix summarizes these interactions.   }
    \label{fig:Figure_1}
\end{figure} 
Over the past decade, there has been a surge in quantitative modeling of the cellular processes and signaling pathways that regulate bone remodeling.  The first such models, developed by Lemaire \cite{lemaire2004modeling}, Komarova \cite{komarova2003mathematical,komarova2005mathematical}, and colleagues, focused on the temporal dynamics of remodeling at a fixed location in the bone. Based on systems of ordinary differential equations (ODE), these models successfully described the interactions between osteoclasts and osteoblasts and the resulting bone mass balance. The original ODE models have since been applied and extended by various authors, see e.g.\ the work by Pivonka, \cite{pivonka2008model}, Buenzli \cite{buenzli2012modelling}, Ji \cite{ji2012novel}, and colleagues. For further references, as well as an overview of modeling studies with focus on the mechanical aspects of remodeling, we refer to the review articles \cite{pivonka2010mathematical,webster2011silico}.\\

While ODE models provide valuable insights into the complex dynamics of physiological and pathological bone turnover, they are not able to capture salient spatial features of the remodeling process \cite{parfitt1994osteonal}. In fact, the latter takes place on the complex geometries of cortical and trabecular bone, and paracrine signaling between bone cells, which is mediated by soluble chemokines, allows for non-local regulation \cite{khosla2001minireview}.  To model such non-local phenomena, our group \cite{ryser2008mathematical,ryser2010cellular,ryser2012osteoprotegerin} and others \cite{ buenzli2011spatio,graham2012towards} previously developed partial differential equation (PDE) models of bone remodeling. In addition, discrete agent-based models of the spatial remodeling dynamics  were introduced to study the dynamics of individual remodeling units \cite{buenzli2012investigation,van2008unified,van2011simulations}. Spatial aspects of the remodeling biology are also captured in various biomechanical models of bone adaptation   \cite{badilatti2016large,cox2011analysis,scheiner2013coupling}.  \\

These spatial extensions of the original ODE models are endowed with high-dimensional parameter spaces and their analyses rely on computer simulations. In consequence, to gain mechanistic insights and understand which model components are relevant to regulate and maintain physiological remodeling, systematic and extensive parameter space explorations are necessary, and a complete characterization of the dynamic regimes is generally beyond reach. Furthermore, most spatial models focus on osteoclast and osteoblast dynamics only, while treating bone as a passive constituent that is either resorbed and deposited by the two active players of the process. Based on experimental evidence \cite{nakashima2011evidence,xiong2011matrix} however, it has become clear that quiescent bone and embedded, mechanotransducing osteocytes play a key role in the regulation of remodeling.\\

In view of the above limitations of current spatial models, our objective was to develop a spatial model of bone remodeling biology that (i) is amenable to analysis and complete classification in the sense of the original, reductionist ODE model by Komarova and colleagues \cite{komarova2003mathematical}; and (ii) treats quiescent bone and embedded osteocytes as an active part of the remodeling dynamics. We  focused on trabecular remodeling and developed our model in the framework of evolutionary game theory (EGT). The latter was introduced by Maynard Smith in 1982 \cite{smith1982evolution}, and has since been used to study a wide range of systems in biology and ecology  \cite{frey2010evolutionary,hammerstein1994game,nowak2006evolutionary,broom2013game}. The analysis of spatial EGT models poses substantial technical difficulties and is a field of active research. Recent advances by Cox, Durrett and Perkins \cite{cox2013voter} and Durrett  \cite{durrett2014spatial} on the weak selection limit for EGTs enabled the analyses in this article.

%%%%%%%%%%%%%%%%%%%%%%%%%%
%%%%%%%%%%%%%%%%%%%%%%%%%%
\section{Methods}\label{model}
%%%%%%%%%%%%%%%%%%%%%%%%%%
%%%%%%%%%%%%%%%%%%%%%%%%%%

%%%%%%%%%%%%%%%%%%%%%%%%%%
\subsection{Spatial Model}\label{biology}
%%%%%%%%%%%%%%%%%%%%%%%%%%

We start by introducing the general idea, and then proceed to construct the formal process. To model physiological remodeling of trabecular bone (Figure \ref{fig:Figure_1}A) in a discrete spatial setting, we partition the volume of trabecular bone into zones of bone resorption, bone formation and quiescent bone. Zones of resorption are populated by bone matrix degrading osteoclasts, and zones of formation are  populated by osteoid producing osteoblasts. Quiescent zones on the other hand consist of bone matrix and embedded osteocytes. We then allow the different zones to interact in a probabilistic manner, resulting in growing and shrinking patches of resorption, formation and quiescence. For example, if a zone of formation (osteoblasts) is adjacent to a zone of quiescence (bone), then the zone of formation is expected to  convert to a zone of quiescence, consisting of newly formed bone with embedded osteocytes. Conversely, if a zone of quiescence (bone) is adjacent to a zone of resorption (osteoclasts), the former is expected to vanish and be replaced by the expanding zone of resorption. The resulting process is an evolutionary competition between neighboring zones. Due to the complex interactions between cell types, the probability of each zone to invade or to be invaded depends on the make-up of its neighborhood.\\

To formally construct this spatial evolutionary process, we consider a fixed bone volume and partition it into a regular three-dimensional lattice with $N$ elements. Each element is occupied by one of the three zones, and the zones are labeled as type $1$ (resorption), type $2$ (formation) and type $3$ (quiescence). There is flexibility with respect to the physical size attributed to the lattice elements. However, the side length of each element needs to be larger than the size of  individual osteoclasts because they are the largest bone cells and measure approximately $50$ microns in diameter \cite{eriksen1992cellular}. In addition, the lattice elements should be small enough to allow for sufficient spatial resolution of the process.\\

We denote by $\xi_t(x)\in \{1,2,3\}$ the type of zone occupying lattice element $x$ at time $t$. Following the basic principles of EGT, see also \cite{durrett2014spatial}, the stochastic expansion rate $\psi_t(x)$ of element $x$ at time $t$ (referred to as its {\it fitness} in EGT), is determined by the make-up of its surrounding elements:
\begin{align}\label{e0}
\psi_t(x)=\sum_{y\sim x} \bar G (\xi_t(x),\xi_t(y)),
\end{align}
where $y\sim x$ denotes the set of nearest neighbors of $x$, and $\bar G$ is the so-called {\it pay-off matrix} of the bone remodeling game, see Figure \ref{fig:Figure_1}B. More precisely, $\bar G(i,j)$, also denoted as $\bar g_{ij}$, is the expansion rate of a type-$i$ zone in the presence of a type-$j$ zone, see Figure \ref{fig:Figure_1}C. For example, $\bar g_{12}$ is the expansion rate of a resorption zone in presence of a formation zone, and $\bar g_{31}$ is the expansion rate conferred to a quiescent zone by a formation zone. According to its instantaneous expansion rate $\psi_t(x)$, element $x$ will stochastically expand and replace one of its nearest neighbors, chosen uniformly at random. Since each element has its own expansion rate -- which depends on the constituency of its neighbors through \eqref{e0} -- this defines a global stochastic process on the lattice:  elements with higher expansion rates tend to take over their neighbors faster, and less proliferative elements in turn are eliminated by their expanding neighbors. Finally, to be consistent with the interpretation of an expansion rate, the pay-off matrix $\bar G$ is assumed to be positive. For reasons that will become clear below, we rewrite $\bar G\equiv {\bf 1}+ \omega G,$ where ${\bf 1}$ is the $3\times3$ matrix consisting of all $1$'s, $G=(g_{ij})$ is a pay-off matrix with real-valued and possibly negative entries, and $\omega>0$ is chosen small enough to ensure positivity of $\bar G$. 

%%%%%%%%%%%%%%%%%%%%%%%%%%
\subsection{ Parameter Considerations}
%%%%%%%%%%%%%%%%%%%%%%%%%%
Taking into consideration established knowledge about the biology of bone remodeling, we can make {\it a priori} restrictions on the 9-dimensional parameter space defined by the pay-off matrix $G$. These constraints are summarized and justified in Table 1.

\begin{table}[htb!]
\centering
\begin{tabular}{llp{8cm}ll}\label{Table1}
 Parameter & Constraint  & Justification & References   \\
 \hline
$g_{11}$ & $>0$  &  Autocrine stimulation by TNF-$\alpha$, IL-1$\alpha$ & \cite{zou2001tumor,tani1999autocrine}  \\
$g_{12}$  & $\in \mathbb R$ & Net impact depends on RANKL/OPG balance   & \cite{khosla2001minireview,quinn2001transforming}    \\
 $g_{13}$& $>0$  & Release of matrix-embedded growth factors;  osteocyte-derived RANKL stimulation   &  \cite{nakashima2011evidence,xiong2011matrix,teitelbaum2000bone}\\
 $g_{21}$ & $>0$ & Paracrine stimulation by TGF-$\beta$, IGF & \cite{erlebacher1998osteoblastic, canalis1996insulin} \\ 
 $g_{22}$ & $= 0$ & Presumed negligible & \cite{ryser2008mathematical} \\
 $g_{23}$ & $<0$ &  Wnt-Sclerostin signaling, considered inhibitory. & \cite{li2005sclerostin,winkler2003osteocyte} \\
 $g_{31}$ & $<0$ & Osteoclasts resorb bone & \cite{teitelbaum2000bone} \\
 $g_{32}$ & $>0$ & Osteoblasts produce bone matrix and become embedded osteoctyes & \cite{ducy2000osteoblast} \\
 $g_{33}$ & $=0$ & Osteocytes are terminally differentiated and do not produce or resorb bone & \cite{bonewald2007osteocytes} \\
 \hline
  $\alpha_1$ & $<0$ & $\alpha_1=g_{23}$ & -\\
  $\alpha_2$ & $<0$ & $\alpha_2=g_{31}-g_{11}$ & -\\
     $\alpha_3$ & $\in \R$ &  $\alpha_3=g_{12}$ & -\\
  $\beta_1$ & $>0$ & $\beta_1=g_{32}$ & -\\
   $\beta_2$ & $>0$ & $\beta_2=g_{13}$ & -\\
   $\beta_3$ & $\in \R$ & $\beta_3=g_{21}-g_{11}$ & -\\
\end{tabular}
\caption{{\bf Parameter Constraints.} {\it Top:} Summary of the a priori constraints on the model parameters $g_{ij}$ based on published biological findings. $g_{ij}$ quantifies the impact of a zone of type $j$ on a zone of type $i$. {\it Bottom:} The resulting constraints for $\alpha_i$ and $\beta_i$.  }
\end{table}Furthermore, because subtracting $g_{11}$ from the first column in $G$affects neither the non-spatial replicator dynamics nor the weak selection limit of the spatial game \cite{durrett2014spatial}, we will henceforth consider the transformed matrix $G$,
\begin{align}\label{Gmatrix1}
G=\left( \begin{array}{ccc}
0 & \alpha_3 & \beta_2 \\
\beta_3 & 0 & \alpha_1 \\
\alpha_2 & \beta_1 & 0
 \end{array} \right), \quad \alpha_1, \alpha_2<0,  \qquad \beta_1, \beta_2>0, \qquad \alpha_3, \beta_3 \in \R,
 \end{align} 
see Table 1 for the definitions of  $\alpha_i$ and $\beta_i$ in terms of the $g_{ij}$.

%%%%%%%%%%%%%%%%%%%%%%%%%%
\subsection{ Replicator Dynamics of Non-spatial Model}
%%%%%%%%%%%%%%%%%%%%%%%%%%
Spatial models are more difficult to analyze than their temporal counterparts and analyses rely on extensive simulations. Therefore, an important question in every spatial modeling study is the necessity to account for space explicitly. To address this issue, we introduce here the non-spatial version of the evolutionary game model. This version neglects spatial structure and assumes that the distinct trabecular zones are well-mixed and thus all equally likely to interact with each other. Instead of analyzing the fully stochastic system, we notice that the number of lattice elements in a bone is large (assuming a diameter of $50-100$ microns), and hence we can study the problem in the deterministic limit as $N\to\infty$. In this approximation, the non-spatial EGT dynamics are described by the standard replicator dynamics from EGT \cite{smith1982evolution}. Formally, we denote by $x(t):=(x_1(t), x_2(t), x_3(t))\geq (0,0,0)$ the densities of resorptive ($x_1$), formative ($x_2$) and quiescent ($x_3$) zones, respectively, with $x_1+x_2+x_3\equiv 1$ at all times $t\geq 0$. Then, as $N\to \infty$, the non-spatial dynamics of the well-mixed system are described by the replicator equations 
\begin{align}\label{e1}
\dot{x}_i=\phi_G^i(x)\equiv x_i\cb{F_i(x)- \ip{F}\!(x)}, \qquad i=1,2,3,
\end{align} 
where $F_i(x):=(Gx)_i$ is the expansion rate of species $i$, and $\ip{F}\!(x):=x^TGx$ is the average expansion rate of the entire population \cite{smith1982evolution}. The interior fixed point for the replicator dynamics (\ref{e1}), if it exists, is given by
\begin{align}\label{fpoints}\begin{split}
\rho_1= & (\beta_1\beta_2+\alpha_1\alpha_3-\alpha_1\beta_1)/D\\
\rho_2= & (\beta_2\beta_3+\alpha_1\alpha_2 -\alpha_2\beta_2)/D\\
\rho_3= & (\beta_1\beta_3+\alpha_2\alpha_3-\alpha_3\beta_3)/D,\end{split}
\end{align}
 where $D$ is the sum of the three numerators. The dynamics of the replicator equation \eqref{e1} are discussed in Appendix \ref{app_mean}.
 
 %%%%%%%%%%%%%%%%%%%%%%%%%%
\subsection{ Numerics}
%%%%%%%%%%%%%%%%%%%%%%%%%%

All model simulations were performed using the software MatLab (Version 8.5.0, The MathWorks Inc.\ 2015). The built-in Runge-Kutta solver {\it ode45} was used to solve the deterministic replicator equations. For the fully spatial model, we used a Gillespie algorithm to simulate the stochastic process on a cubic lattice with $L^3$ nodes and periodic boundary conditions.  Initial fields were generated using a product measure as specified in the figure captions.
%%%%%%%%%%%%%%%%%%%%%%%%%%
\section{Results}\label{results}
%%%%%%%%%%%%%%%%%%%%%%%%%%

In this section, we first analyze the fully spatial model, classify its dynamic regimes and identify zones in parameter space that allow for coexistence of resorption, formation and quiescence. We then study emerging spatial structure based on three-dimensional simulations of the evolutionary process. Finally, we compare the non-spatial and spatial versions of the game.

%%%%%%%%%%%%%%%%%%%%%%%%%%
\subsection{The Spatial Game}\label{spat_game}
%%%%%%%%%%%%%%%%%%%%%%%%%%

%There are several possibilities for rendering the replicator dynamics spatial. Here, we consider the spatial dynamics as a perturbation of the voter model dynamics. This choice is motivated by the fact that the theory of such perturbed voter models has been developed substantially over the past years, allowing for analytic insight into the resulting dynamics. Similar results are not available for other spatial versions of evolutionary games. We will come back to alternative models in the discussion. Let us now explain the perturbed voter dynamics in detail. First, we define the matrix   
%$$\bar G={\bf 1} + \omega G,$$ where ${\bf 1}$ is the matrix of all 1's, $\omega>0$ is the  perturbation parameter and  $G$ is the game matrix introduced above. Next, we think of a three-dimensional lattice where each element is occupied by either a zone of resorption, formation or quiescence. We consider now the following birth-death dynamics: the occupant of site $y$ creates an identical copy of itself at rate $\nu(y)$, and replaces a neighbor (chosen uniformly at random) with its copy. Importantly, the proliferation rate is defines as the fitness of $y$ with respect to its nearest neighbors:
%$$\nu(y):= \sum_{z\sim y} \bar G(\xi(y),\xi(z)),$$ where $\sim$ denotes nearest neighbor, and $\xi(y)$ is a function that identifies the type of zone $y$. 

\begin{figure}[htb!] 
  \centering
   \includegraphics[width=1\textwidth]{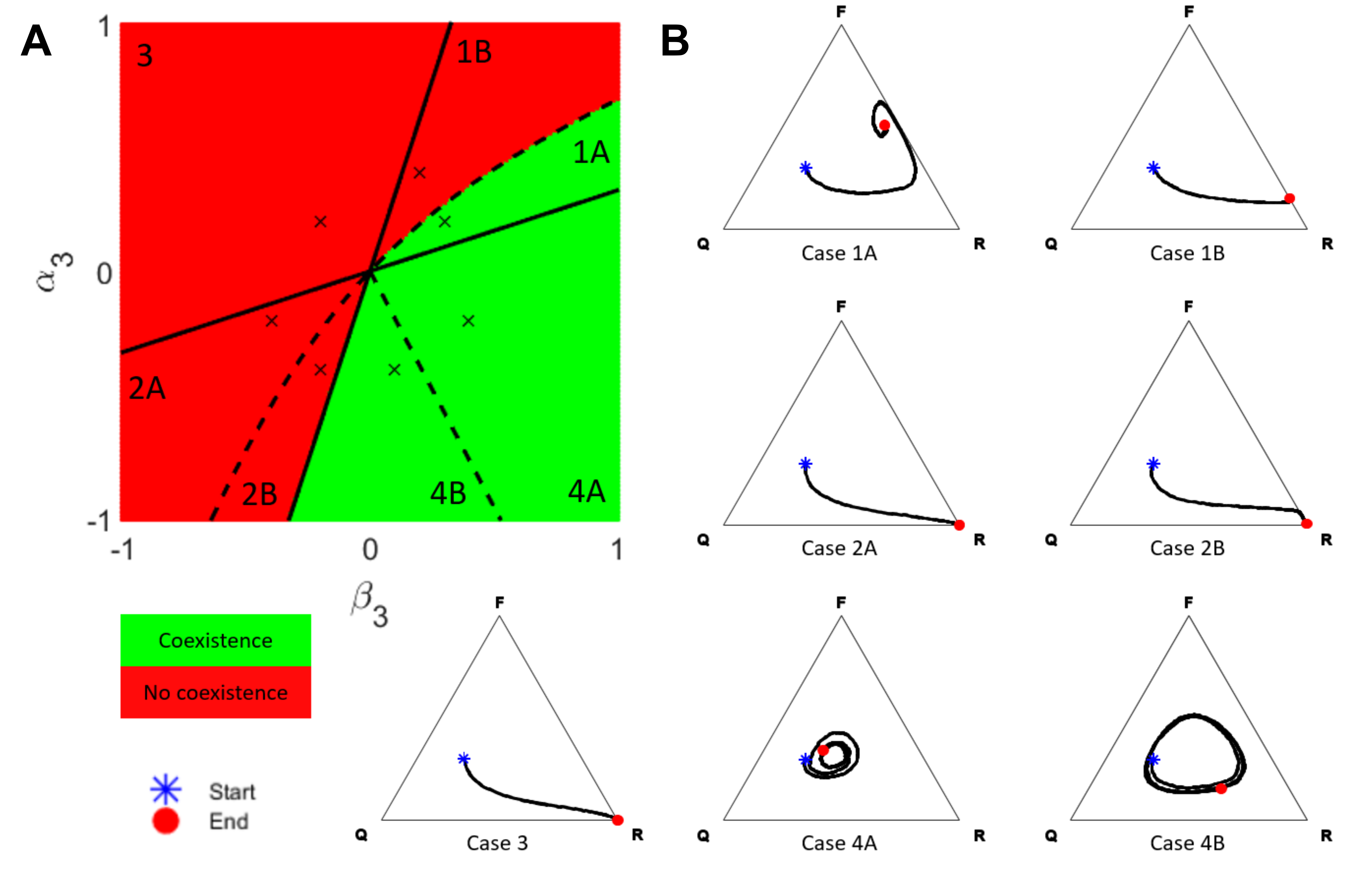} 
   \vspace{-.4cm}
   \caption{ \footnotesize  {\bf Coexistence in the Spatial Model.} {\bf (A)}  Phase diagram showing regions of coexistence (green) and lack of coexistence (red) in the  ($\alpha_3,\beta_3$)-plane. The ($\times$) mark the parameter choices for the examples in panel B; the remaining model parameters were set to $\alpha_1 = -0.1, \alpha_2 = -0.5, \beta_1 = 0.6, \beta_2 = 0.2, \theta = 0.485$, $\omega=0.1$. {\bf (B)} For each case in panel A, a realization of the spatial stochastic process is simulated on a cubic lattice of side length $L=100$ over $2\cdot 10^9$ iterations (corresponding to $\approx 2000$ time units). At simulation {\it Start}, the field is seeded using a product measure with probabilities 0.2 (R), 0.3 (F) and 0.5 (Q), respectively.  The resulting trajectories are visualized in ternary plots and terminate at the {\it End} symbol. The values of $\alpha_3$ and $\beta_3$ are indicated by ($\times$) in panel A, and all remaining parameters as specified above. }
     \label{fig:Figure_3}
\end{figure}

The spatial game dynamics are determined by the fitness function $\psi_t(x)$ in equation \eqref{e0}, with transformed pay-off matrix $\bar{G}={\bf 1} + \omega G$, where $G$ is the pay-off matrix \eqref{Gmatrix1} and $\omega>0$ is small enough so that all entries of $\bar G$ are  positive. The resulting dynamics are a perturbation of the well-studied voter model, see e.g.\! \cite{liggett2013stochastic}. Thanks to recent theoretical results by Cox, Durrett and Perkins \cite{cox2013voter} and Durrett \cite{durrett2014spatial}, the behavior of the spatial stochastic model can be analyzed in  the weak selection limit. More precisely, if we let $\omega \to 0$ and simultaneously shrink space by $\sim \! \omega^2$ and speed up time by $\sim \! 1/\omega$, then the temporal evolution of the density of species $i$ at location $x$, denoted by $u_i(x,t)$, evolves according to the PDE 
\begin{align}\label{birthdeath}
\frac{\partial u_i(x,t)}{\partial t}=\frac{1}{6}\Delta u_i(x,t) + \phi_H^i(u(x,t)).
\end{align}
Here, $u=(u_1,u_2,u_3)$ and $\phi_H^i$ is the rate of change on the right-hand side of the replicator equation \eqref{e1}, with the pay-off matrix $G$ replaced by $H$ defined as \cite{durrett2014spatial}
$$H_{ij}=G_{ij}+\theta \ob{G_{ii}+G_{ij}-G_{ji}-G_{jj}}. $$
The constant $\theta$ cannot be calculated exactly, but numerical simulations estimate  $\theta\approx 0.485$ \cite{durrett2014spatial}. In terms of $\alpha_i$ and $\beta_i$, the explicit expression of the $H$ matrix is
%$$ \phi_H^i(u)= p_1 \phi_R^i(u)+ p_2 \sum_{j\neq i} u_i u_j (G_{ii}+G_{ij}-G_{ji}-G_{jj}),$$
 %and $\phi_R^i(u)$ is the rate of change on the right-hand side of the replicator equation  \eqref{e1}. The constants $p_1$ and $p_2$ cannot be exactly calculated, but numerical bounds are given by $p_1 \in [0.32, 0.33]$ and $p_2 \in [0.155, 0.161]$, see  \cite{durrett2014spatial}. 
%=p_1/p_2$,   $p_1\approx 0.32$, and $p_2\approx 0.16$ according to numerical simulations
%In addition, one can show that $\phi_H^i$ is the right-hand side of the replicator equation \eqref{e1} where the payoff matrix $G$ is replaced by the transformed pay-off matrix \cite{durrett2014spatial} $$H_{ij}=G_{ij}+\theta \ob{G_{ii}+G_{ij}-G_{ji}-G_{jj}},\qquad \theta=p_2/p_1\approx 1/2. $$ In terms of $\alpha_i$ and $\beta_i$, the explicit expression of the $H$ matrix is
\begin{align}\label{Hmatrix}
H=\left( \begin{array}{ccc}
0 & (1+\theta)\alpha_3 -\theta \beta_3& (1+\theta)\beta_2 - \theta \alpha_2 \\
(1+\theta)\beta_3-\theta \alpha_3 & 0 & (1+\theta)\alpha_1 -\theta \beta_1 \\
(1+\theta) \alpha_2 -\theta \beta_2 & (1+\theta)\beta_1 -\theta \alpha_1 & 0
 \end{array} \right), 
 \end{align} 
with the following constraints imposed by \eqref{Gmatrix1}: $h_{13}, h_{32}>0$, $h_{23}, h_{31}<0,$ and $h_{12}, h_{21} \in \R$. In addition to the limiting behavior of the PDE \eqref{birthdeath}, there are analytic coexistence results for $\omega$ finite but small enough \cite{cox2013voter}. Details of the complete analysis are found in Appendix \ref{app_spatial}, and the results are summarized in Figure \ref{fig:Figure_3}A,
where the phase diagram is projected onto the $(\alpha_3,\beta_3)$-plane, distinguishing zones of three-species coexistence (green) and lack of coexistence (red). Because we were not able to determine the qualitative behavior of all 7 scenarios based on theoretical results, we ran representative three-dimensional simulations to confirm the conjectured behavior in each case, see Figure \ref{fig:Figure_3}B.\\

\begin{figure}[htb!] 
  \centering
   \includegraphics[width=1\textwidth]{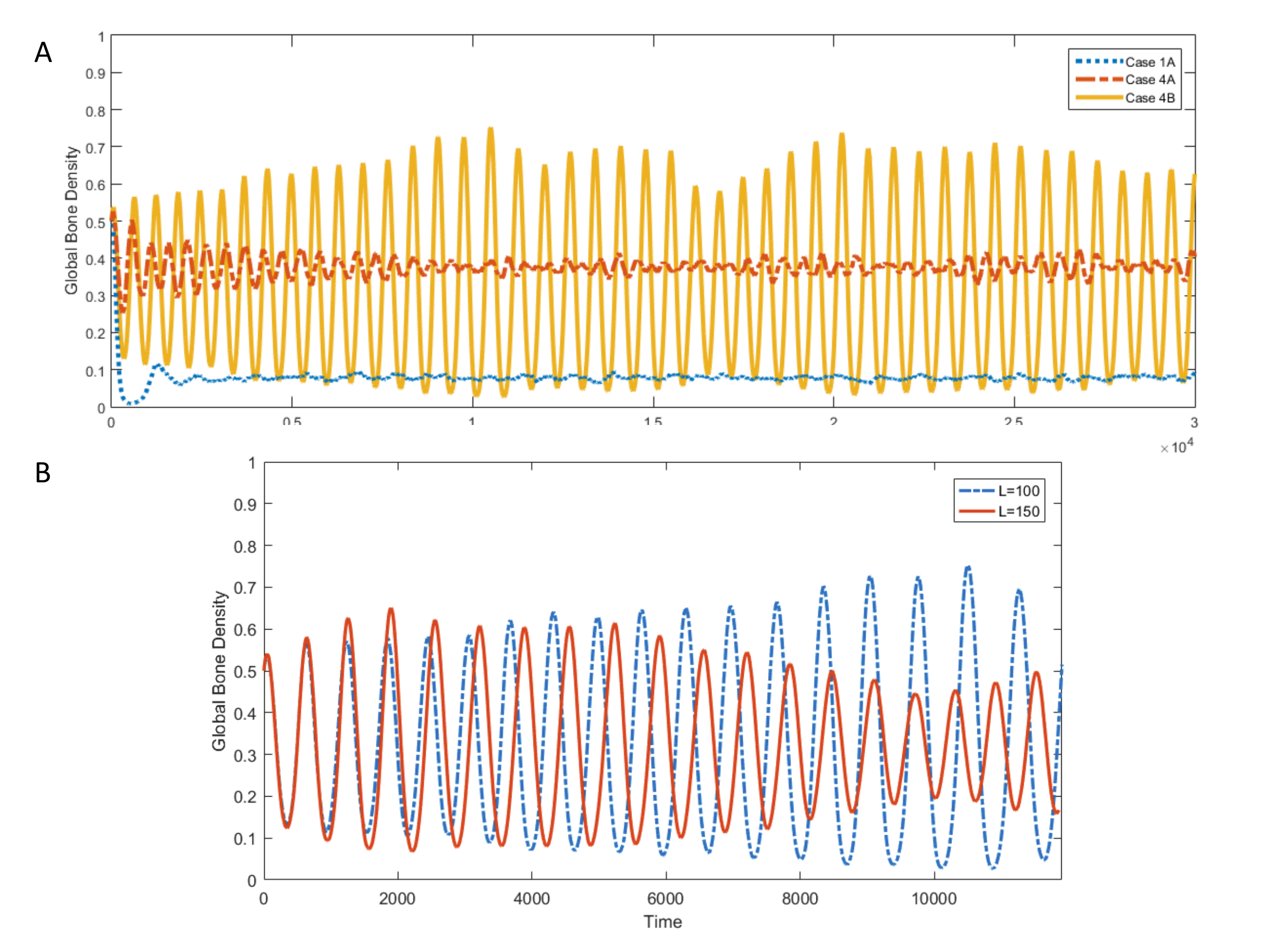} 
   \vspace{-.4cm}
   \caption{ \footnotesize  {\bf Bone Density Evolution in Coexistence Regime.}  {\bf (A)} For Cases 1A, 4A and 4B, which all exhibit long-term coexistence (see Figure \ref{fig:Figure_3}A), the evolution of the global bone density is shown over time. In Case 1A, the bone density quickly approaches its steady state value. In Case 4A and 4B, the system undergoes bounded oscillations. In all simulations, $\theta=0.485$, $\omega=0.1$, and remaining parameters as follows: Case 1A $\alpha_1 = -0.1, \alpha_2 = -0.5, \alpha_3 = 0.2, \beta_1 = 0.6, \beta_2 = 0.2, \beta_3 = 0.3$; Case 4A: $\alpha_1 = -0.1, \alpha_2 = -0.5, \alpha_3 = -0.2, \beta_1 = 0.6, \beta_2 = 0.2, \beta_3 = 0.4$; Case 4B: $\alpha_1 = -0.1, \alpha_2 = -0.5, \alpha_3 = -0.4, \beta_1 = 0.6, \beta_2 = 0.2, \beta_3 = 0.1$. {\bf (B)} For Case 4B, the process is simulated for two domain sizes, $L=100$ and $L=150$, respectively. Remaining parameters: $\theta=0.485$, $\omega=0.1$, $\alpha_1 = -0.1, \alpha_2 = -0.5, \alpha_3 = -0.4, \beta_1 = 0.6, \beta_2 = 0.2, \beta_3 = 0.1$.}
     \label{fig:Figure_3_new}
\end{figure}

As illustrated in  Figure \ref{fig:Figure_3}A, the phase transition from no coexistence (red) to coexistence (green) occurs across the boundaries between Cases 2B and 4A, and Cases 1A and 1B, respectively. The slope of the linear boundary between Cases 2B and 4A is uniquely determined by the parameters $\alpha_3$ (osteoblast-derived regulation of osteoclasts) and $\beta_3$ (difference between osteoclast-derived stimulation of osteoblasts and osteoclast-derived autocrine stimulation). In contrast, the nonlinear shape of the boundary between Cases 1A and 1B depends on all model parameters, see Appendix \ref{app_spatial} for details. In the regions of no coexistence, there is either complete takeover by resorption (Cases 2A, 2B and 3) or the system converges to a stable resorption-formation equilibrium (Case 1B), see Figure \ref{fig:Figure_3}B. Among cases with three-species coexistence, Case 1A converges to a fixed point in the interior of the simplex, whereas Cases 4A and 4B exhibit irregular oscillations that are bounded away from the edges of the simplex, see Figure \ref{fig:Figure_3_new}A. Increasing the domain size was found to attenuate the oscillations in Case 4B, see Figure \ref{fig:Figure_3_new}B.

%The qualitative behavior of the spatial process is primarily determined by two parameters: $\alpha_3$, the strength of osteblast-derived regulation of osteoclasts, and $\beta_3$, the difference between osteoclast-derived stimulation of osteoblasts and osteoclast-derived autocrine stimulation, see Appendix \ref{app_spatial}. The remaining four interaction parameters $\alpha_1$, $\alpha_2$, $\beta_1$ and $\beta_2$ only affect the shapes of the boundaries  between cases 1A/1B, 2A/2B and 4A/4B, see the dotted lines in Figure \ref{fig:Figure_3}A. Importantly, the only such boundary that corresponds to a phase transition is the transition between 1A (no coexistence) and 1B (coexistence). In contrast, the transition between 2A (no coexistence) and 2B (coexistence) is independent of all other interaction parameters. Among cases without coexistence, there is either complete takeover by resorption (Cases 2A, 2B and 3) or the system converges to a stable resorption-formation equilibrium (Case 1B), see Figure \ref{fig:Figure_3}B. On the other hand, among cases with three-species coexistence, Cases 1A and 4A converge toward an interior fixed point, whereas Case 4B appears to approach a limit cycle, see Figure \ref{fig:Figure_3_new}.

%%%%%%%%%%%%%%%%%%%%%%%%%%
\subsection{Emerging Spatial Structure}
\begin{figure}[htb!] 
  \centering
   \includegraphics[width=.9\textwidth]{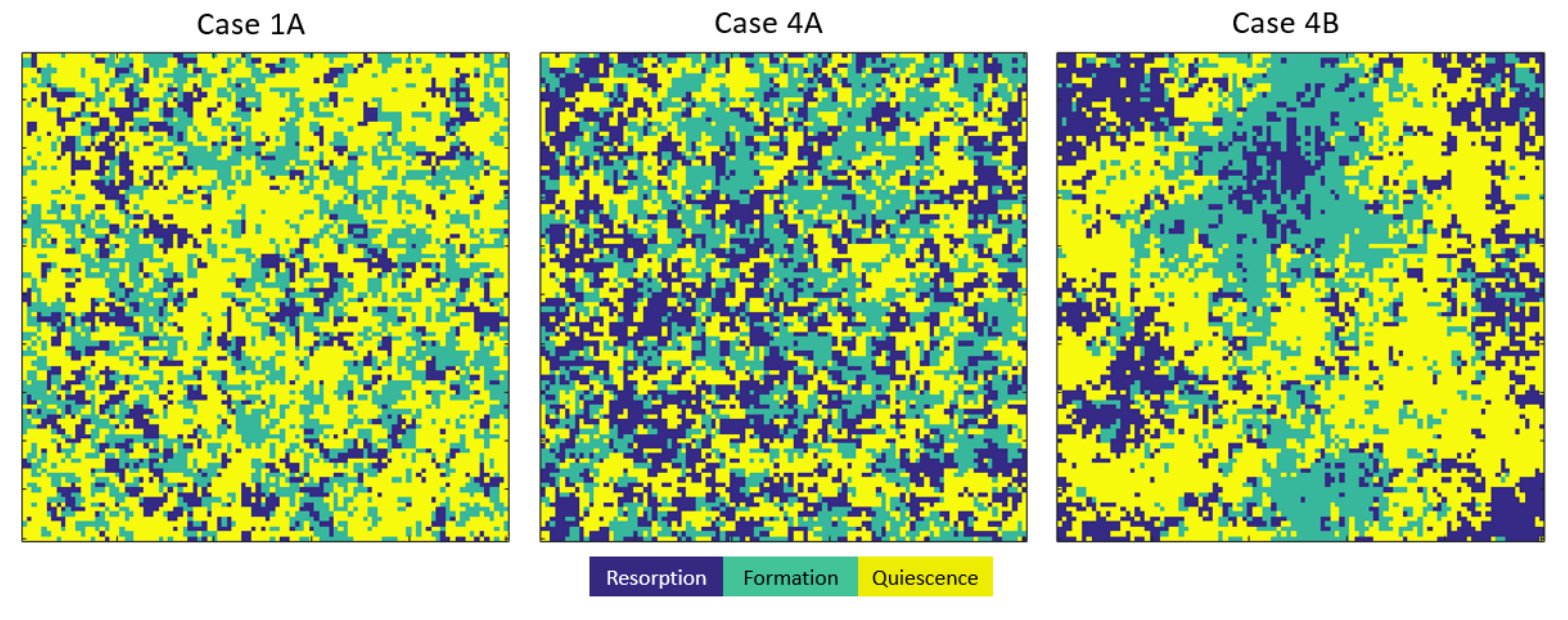} 
   \vspace{-.4cm}
   \caption{ \footnotesize  {\bf Spatial Structure of Coexistence.} For each of the three cases that allow for coexistence in the spatial game (see Figure \ref{fig:Figure_3}A), a realization of the  stochastic process was simulated on a cubic lattice of side length $L=100$. The initial field is seeded using a product measure with probabilities 0.2 (R), 0.3 (F) and 0.5 (Q), respectively, and representative 2D cross sections of the 3D systems are shown after reaching the stationary state. All remaining model parameters as follows. Case 1A $\alpha_1 = -0.2, \alpha_2 = -0.1, \alpha_3 = 0.45, \beta_1 = 0.5, \beta_2 = 0.1, \beta_3 = 0.6$; Case 4A: $\alpha_1 = -0.1, \alpha_2 = -0.5, \alpha_3 = -0.3, \beta_1 = 0.6, \beta_2 = 0.3, \beta_3 = 0.4$; Case 4B: $\alpha_1 = -0.6, \alpha_2 = -0.6, \alpha_3 = -0.4, \beta_1 = 0.6, \beta_2 = 0.2, \beta_3 = 0.4$.}
     \label{fig:Figure_4}
\end{figure}

%%%%%%%%%%%%%%%%%%%%%%%%%%
\noindent Next we sought to characterize the spatial structure of the evolutionary game in regimes that allow for three-species coexistence, i.e.\ Cases 1A, 4A and 4B. To this end, we performed stochastic simulations of the three-dimensional evolutionary process and visualized 2D sections as shown in  Figure \ref{fig:Figure_4}.
Starting from randomly distributed initial conditions, we observed the emergence of spatial clustering of the three coexisting species. The resulting clusters of bone tissue, surrounded by zones of formation and resorption, are reminiscent of the sponge-like patterning in vertebrate trabecular bone. \\

\begin{figure}[htb!] 
  \centering
   \includegraphics[width=.9\textwidth]{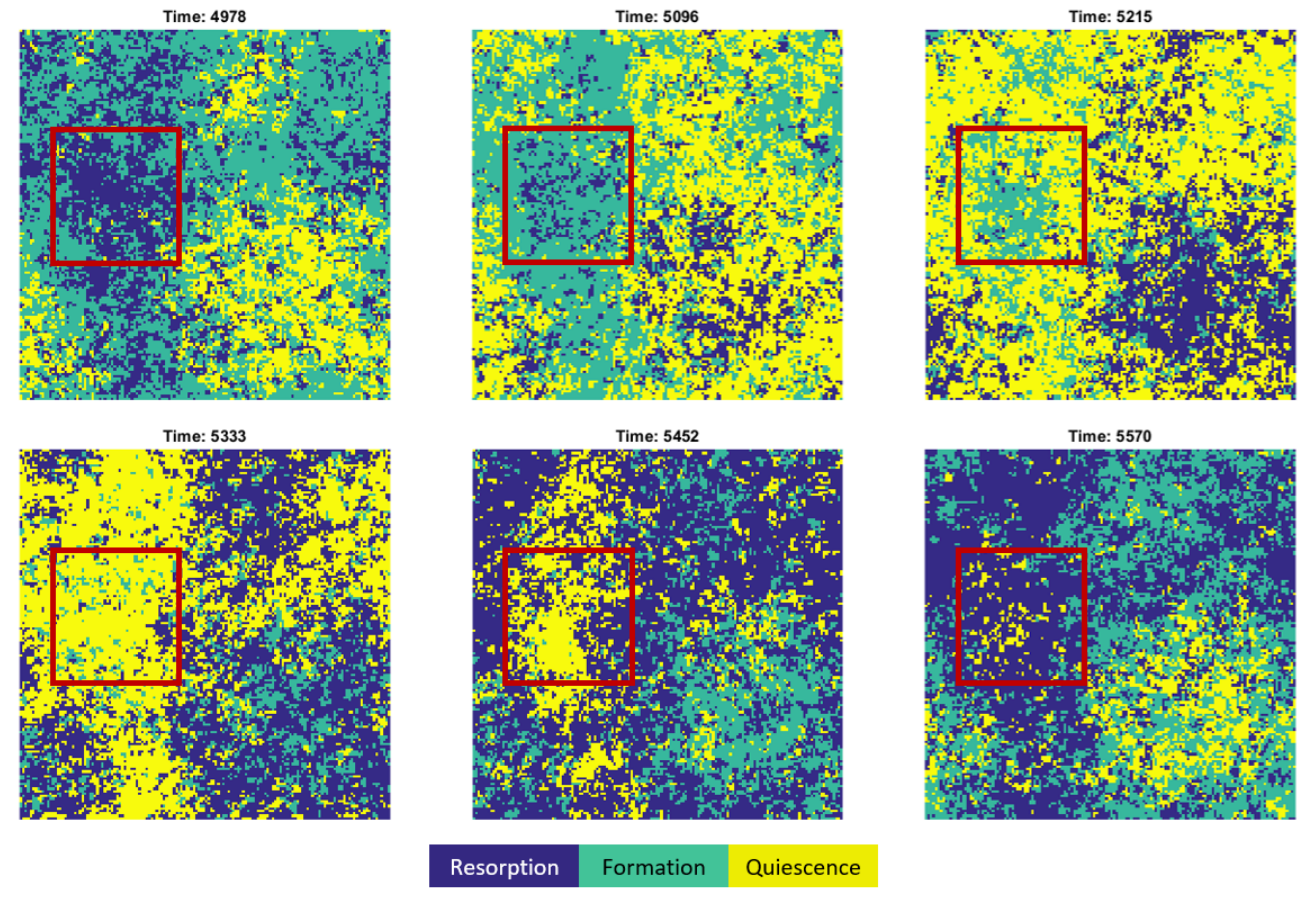} 
   \vspace{-.4cm}
   \caption{ \footnotesize  {\bf The Local Dynamics of Case 4B.} Successive 2D sections of the unstable rock-paper-scissors game (Case 4B, see Figure \ref{fig:Figure_3}A) are shown. A realization of the  stochastic process was simulated on a cubic lattice of side length $L=150$, with initial field seeded using a product measure with probabilities 0.2 (R), 0.3 (F) and 0.5 (Q). Identical 2D sections of the 3D domain are shown at times 4978, 5096, 5215, 5333, 5452, and 5570, respectively. Locally, see red frame, the system is cyclicly dominated as it transitions from primarily resorptive (time 4978) to primarily bone forming (time 5096), to primarily quiescent (time 5333) and back to primarily resorptive (time 5570). Because the total volume consists of asynchronously cycling patches, the global dynamics remain bounded, see also Figures \ref{fig:Figure_3}B and \ref{fig:Figure_3_new}.  Remaining parameter values:  $\theta=0.485$, $\omega=0.1$, $\alpha_1 = -0.6, \alpha_2 = -0.6, \alpha_3 = -0.4, \beta_1 = 0.6, \beta_2 = 0.2, \beta_3 = 0.4$.}
     \label{fig:Figure_5}
\end{figure}

In Section \ref{spat_game} we saw that Case 1A  reaches a global equilibrium state, whereas Cases 4A and 4B exhibit long-time oscillatory behavior, see also Figure \ref{fig:Figure_3_new}. Case 4B is particularly interesting with respect to emerging spatial structure. In fact,  following a fixed volume within the same bone section over time, see the red frame in Figure \ref{fig:Figure_5},
 we found that the patch was cyclicly dominated by formation, quiescence and resorption. As we will see in Section \ref{rol_spac}, these local dynamics are consistent with the outward spiraling trajectories of the associated replicator dynamics. However, because the total volume consists of many such asynchronously cycling patches, the global dynamics become stabilized and exhibit the bounded behavior shown in Figure \ref{fig:Figure_3_new}. Overall, this example illustrates how spatial structure can fundamentally alter the dynamics, and we turn our attention now to a systematic exploration of the role of spatial structure in bone remodeling.

%%%%%%%%%%%%%%%%%%%%%%%%%%
\subsection{ The Role of Space in Bone Remodeling}\label{rol_spac}
%%%%%%%%%%%%%%%%%%%%%%%%%%
%Coexistence is only possible in Cases 1A and 4A;  in all other regions of parameter space, there is either loss of quiescence (Case 1B), loss of both quiescence and formation (Cases 2 and 3), or the system is unstable and spirals outwards (Case 4B). In the remainder of this section, we discuss and compare conditions for coexistence in the well mixed and spatial games.\\

% In summary, an interior fixed point and hence coexistence is only possible if $\beta_3>0$, and either $\alpha_3>0$ with $ \alpha_2\alpha_3+\beta_1\beta_3-\alpha_3\beta_3>0$ (Case 1A), or $\alpha_3<0$ with $\beta_1\beta_2 \beta_3+\alpha_1\alpha_2\alpha_3>0$ (Case 4A). In all other regions of parameter space, there is either loss of quiescence (Case 1B), loss of both quiescence and formation (Cases 2 and 3), or the system is unstable and spirals outwards (Case 4B). \\

To enable a direct comparison between the spatial game and the replicator dynamics of the non-spatial version, we first need to analyze the replicator dynamics.\\

\noindent {\bf Replicator Dynamics.} Similarly to the spatial case, analysis of the replicator dynamics of the non-spatial game revealed seven dynamic regimes, see Appendix \ref{app_mean} for details.
\begin{figure}[htb!] 
  \centering
   \includegraphics[width=1\textwidth]{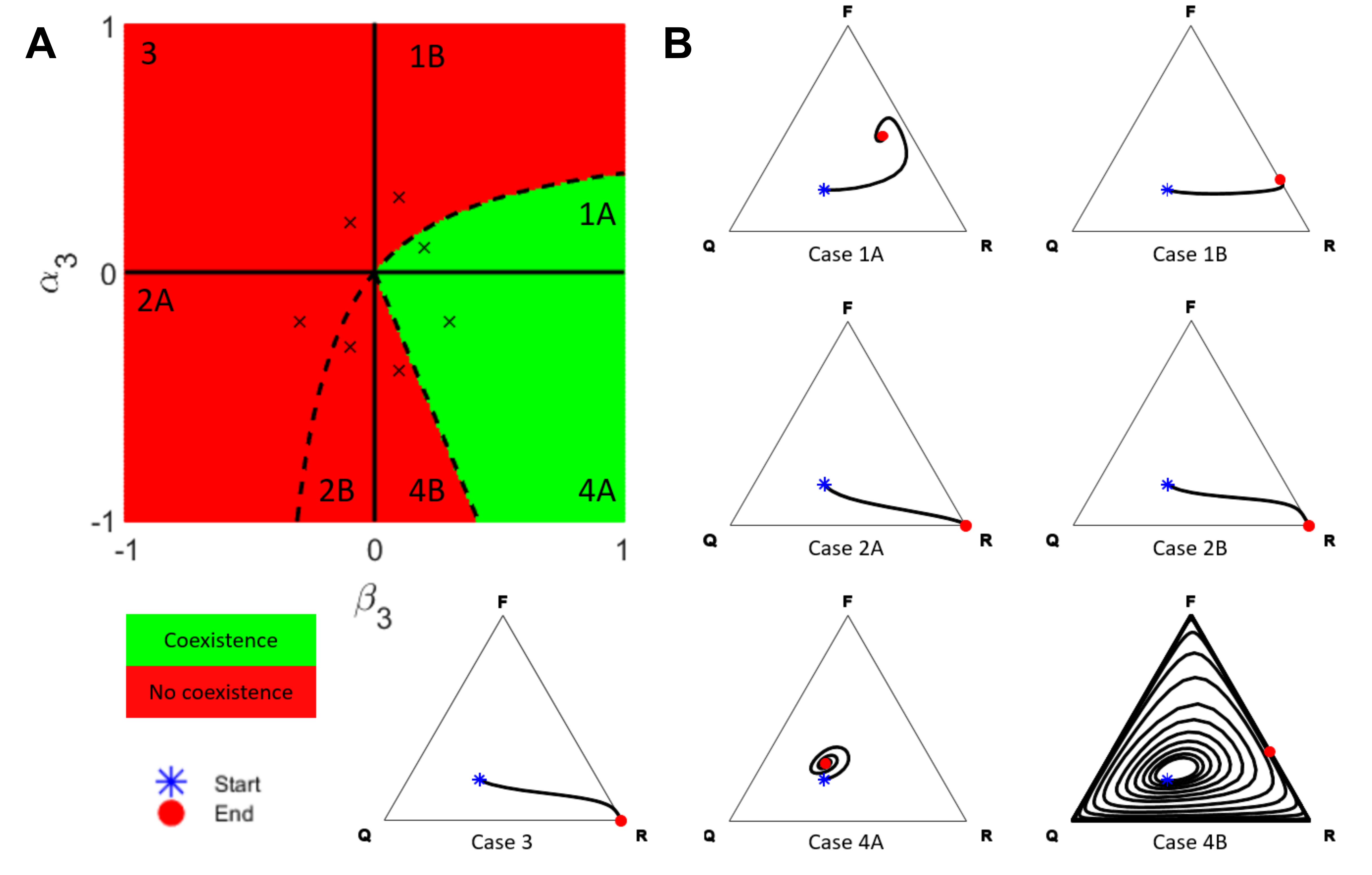} 
   \vspace{-.4cm}
   \caption{ \footnotesize  {\bf Coexistence in the Non-Spatial Model.} {\bf (A)} Phase diagram showing regions of coexistence (green) and lack of coexistence (red) in the  ($\alpha_3,\beta_3$)-plane. The ($\times$) mark the parameter choices for the examples in panel B; the remaining parameters were fixed at  $\alpha_1 = -0.1, \alpha_2 = -0.5, \beta_1 = 0.6, \beta_2 = 0.2$.  {\bf (B)} For each case in panel A, the replicator dynamics are solved up to $t=2000$, with initial conditions $(x_1(0), x_2(0), x_3(0))=(0.3, 0.2, 0.5)$. The values of $\alpha_3$ and $\beta_3$ are indicated by ($\times$) in panel A, and all remaining parameters as specified above. }
     \label{fig:Figure_2}
\end{figure} 
Figure \ref{fig:Figure_2}A illustrates the coexistence regions in a phase diagram projected onto the $(\alpha_3, \beta_3)$-plane, and Figure \ref{fig:Figure_2}B provides a concrete example for each case. As seen in Figure \ref{fig:Figure_2}A, a necessary condition for coexistence is $\beta_3=g_{21}-g_{11}>0$, which means osteoclast-derived paracrine stimulation of osteoblasts ($g_{21}$) dominates osteoclast-derived autocrine stimulation ($g_{11}$). If this condition is not satisfied, i.e.\! $\beta_3<0$, resorption outperforms formation and the global bone density continues to decrease until it vanishes, see Cases 2A, 2B and 3 in Figure \ref{fig:Figure_2}B. On the other hand, if $\beta_3>0$, then coexistence is possible for specific parameter combinations, namely Cases 1A and 4A. The boundaries between physiological and pathological regimes (dashed lines between Cases 1A/1B and 4A/4B, respectively) depend on the values of the remaining model parameters in a nonlinear fashion. These boundaries specify upper and lower bounds for $\alpha_3$  that allow for physiological remodeling, and coexistence is possible for a range of both negative and positive values. In case 4B,  the system cycles through resorption-, formation- and quiescence-dominated regimes and spirals towards the boundary of the simplex, see Figure \ref{fig:Figure_2}B. \\ %Finally, based on the phase diagram in Figure \ref{fig:Figure_2}A, it is seen that the fraction of parameter space allowing for coexistence is bounded by 50\%  from above, and by 0\% from below.\\

\noindent {\bf Comparison to Spatial Dynamics.} Comparing the phase diagram of the spatial game in Figure \ref{fig:Figure_3}A to its non-spatial counterpart in Figure  \ref{fig:Figure_2}A, we make several  observations. First, the four uniform quadrants corresponding to Cases 1, 2, 3 and 4 of the non-spatial game are transformed into two larger (Cases 3 and 4) and two smaller (Cases 1 and 2) sections in the spatial game. It follows that for a given parameter set, the dynamics of the non-spatial version can be fundamentally different from the dynamics of the spatial version. This is particularly striking in Case 4B, which is unstable in the non-spatial setting, but becomes stabilized in the spatial setting: instead of spiraling outward towards the boundary of the simplex in absence of spatial structure (see Case 4B in Figure \ref{fig:Figure_2}B), the system remains confined to the interior of the simplex in the spatial model (see Case 4B in Figures \ref{fig:Figure_3}B and \ref{fig:Figure_3_new}). Due to this stabilizing effect of space, the entire Case 4 allows for coexistence in the spatial setting. In other words, spatial structure stabilizes the dynamics.\\

%%%%%%%%%%%%%%%%%%%%%%%%%%%%%%%%%%%%%
\section{Discussion}\label{discussion}
%%%%%%%%%%%%%%%%%%%%%%%%%%%%%%%%%%%%%

The remodeling of trabecular bone is an intrinsically spatial process regulated by complex cellular and biochemical processes. To date, most mathematical models of the biological and biochemical mechanisms of remodeling have been formulated in non-spatial settings. Existing spatial generalizations of these models suffer from two shortcomings: they are high-dimensional and not amenable to systematic analyses and they do not account for the the role played by bone-embedded osteocytes. In this work, we sought to overcome these limitations by developing a three-dimensional evolutionary game theory model of bone remodeling that explicitly accounts for bone-embedded osteocytes. \\ 

The proposed model describes the nonlinear interactions between zones of resorption, formation and quiescence in a reductionist framework and is amenable to analysis in both the spatial and non-spatial settings. Direct comparison between spatial (Figure \ref{fig:Figure_3}A) and non-spatial (Figure \ref{fig:Figure_2}A) models revealed the existence of parameter space regions that lead to coexistence of resorption, formation and quiescence in the spatial setting, but not in the well-mixed setting (see Case 4B). This emphasizes the critical role of spatial structure in enabling physiological remodeling regimes, and highlights the necessity to use fully spatial models when seeking to elucidate the biological mechanisms of the process.\\

Case 4B, also known as the unstable rock-paper-scissors game \cite{smith1982evolution}, is a particularly interesting scenario. In the non-spatial scenario it was found to be unstable with alternating periods of  resorption-, formation- and quiescence-dominated states, reminiscent of the uncontrolled episodes of resorption and formation in Paget's disease \cite{raisz1999physiology, feng2011disorders}. In the spatial setting on the other hand, cyclic turnover of the three types remained present within spatially separated patches, but due to asynchronous cycling of the patches, the overall dynamics became stabilized. Based on the reasoning by Durrett and Levin  \cite{durrett1998spatial} and simulations (Figure \ref{fig:Figure_3_new}B), we conjecture that the cyclic behavior may be a finite size effect and would eventually disappear for sufficiently large domain sizes. Such properties of non-hierarchical competition models in spatial settings, and the role of space  in dynamic multi-species models in general, have long been acknowledged in the mathematical ecology literature \cite{durrett1998spatial,durrett1994importance}. To our knowledge, we are the first to directly address this issue in the context of bone remodeling.\\

By performing systematic model analyses we identified parameters critical for maintaining physiological remodeling in the sense of three species coexistence. As shown in the phase diagram in Figure \ref{fig:Figure_3}A, two parameters are particularly important for coexistence:  $\beta_3$ $(=g_{21}-g_{11}$), which is the balance between osteoclast-derived stimulation of osteoblasts and osteoclast-derived autocrine stimulation, and $\alpha_3$ ($=g_{12}$), which represents osteoblast-derived regulation of osteoclasts. In the spatial game, coexistence is ensured whenever $\beta_3>0$ and  $\alpha_3<0$, as well as a small extension of this quadrant, see Cases 4A and 4B in Figure  \ref{fig:Figure_3}A. The first constraint, $\beta_3>0$, emphasizes the importance for osteoclasts to effectively recruit osteoblasts after resorption has been completed; deficiencies in this mechanism lead to loss of coexistence due to unbalanced bone resorption. The second constraint, $\alpha_3<0$, requires there to be a negative feedback from osteoblasts to osteoclasts in order to avoid immediate resorption of newly formed tissue. If this constraint is violated, osteoclasts invade zones of formation and trigger onset of pathological remodeling. This is in alignment with our previous findings regarding the critical role of the spatial expression profiles of RANKL and its inhibitor OPG, by which osteoblasts control resorptive activity \cite{ryser2008mathematical,ryser2010cellular,ryser2012osteoprotegerin}. Finally, a note about the role of the remaining model parameters. As long as $\alpha_3$ and $\beta_3$ satisfy the Case 4 constraints, coexistence is guaranteed independent of the values of  $\alpha_1$, $\alpha_2$, $\beta_1$, and $\beta_2$. However, for $\alpha_3$ and $\beta_3$ satisfying the Case 1 constraints, these four parameters determine the shape of the boundary between coexistence (1A) and lack thereof (1B), see dotted line in Figure \ref{fig:Figure_3}A. Most importantly, all six model parameters play a role in determining the quantitative outcome of the spatial game, and hence the bone density in the stationary state of the system. \\

Previously, Dingli and colleagues \cite{dingli2009cancer} used a non-spatial EGT model to study the interactions between multiple myeloma cells, osteoclasts and osteoblasts. Their underlying model of bone remodeling (in absence of multiple myeloma cells) leads to different conclusions, even if analyzed in the spatial context. In fact, it is easy to show that the Dingli model allows for physiological remodeling  only in Cases 1A and 1B of Figure \ref{fig:Figure_2}A (see Section 6 of \cite{durrett2014spatial}). In particular, it exhibits pathological remodeling in the entire lower right quadrant, which was found to exhibit stable coexistence in the three-player game thanks to the presence of osteocyte regulation. While there is insufficient experimental evidence to test these differential predictions, recent experimental \cite{nakashima2011evidence,xiong2011matrix} and theoretical work \cite{graham2013role,buenzli2015osteocytes,buenzli2015quantifying} has emphasized the importance of osteocyte-derived regulation of remodeling. \\

%Roca: \cite{durrett1994importance, roca2009evolutionary},
 %Future work (we do not account for mechanical effects; easy to couple due to the discretization of space and finite element approaches to mechanical modeling; stroma is not modeled explicitly.\\
The model developed in this study captures the dynamics of the cellular interactions between osteoclasts, osteoblasts, and bone-embedded osteocytes in a reductionist, low-dimensional framework. Here, we did not account for mechanical loading, which plays an important role in guiding the overall bone remodeling process \cite{robling2006biomechanical}. However, thanks to the lattice-based formulation of the evolutionary game model, it can easily be coupled to a mechanical loading model, especially within the continuum mechanics framework developed by Hellmich and colleagues \cite{hellmich2004mineral,hellmich2004can}. In this context, it would be  interesting to further investigate the observed clustering dynamics of bone tissue under different regimes of loading, and to characterize the clustering length scales. Finally, the current model constitutes a stepping stone to the study of stromal cells in physiological and pathological remodeling, and the interactions between bone and cancer cells in metastatic bone cancer and multiple myeloma. 

%%%%%%%%%%%%%%%%%%%%%%%%%%%%%%%%%%%%%
\section*{Acknowledgments}
%%%%%%%%%%%%%%%%%%%%%%%%%%%%%%%%%%%%%
The authors are grateful to Prof.\ S.V. Komarova (McGill University) and Prof.\ R. Durrett (Duke University) for fruitful discussions and valuable comments on the manuscript. The authors would like to thank Prof.\ M. Reed (Duke University) and the Math Bio REU program at Duke University for supporting this work. This work was supported by the National Institutes of Health (R01-GM096190); the Swiss National Science Foundation (P300P-154583); and the National Science Foundation (DMS-0943760).

%Here a list of keywords and thoughts we want to try and incorporate in the discussion
%
%\begin{itemize}
%\item lends itself to future mechanical approachs
%\item Mathematical modeling of bone remodeling has seen quite the evolution over the past
%\item Two aspects of bone remodeling that have been neglected by most modelers are space and noise
%\item Fluctuations: many remodeling units, mean-field sense, so it is OK.
%\item However, to use the mean-field equations, and ignore space may not be correct  because we know that space can play a big role in determining coexistence
%\item Describe ode dynamics, for a well-mixed model, then analyze these dynamics ignoring space
%\item Here we took a more detailed approach: stochastic model of pairwise interactions between bone forming, bone resorbing and bone units, analyze well-mixed version (mean field limit)
%\item Framework we use is: evolutionary game theory; motivation is 3-fold.
%\item First, it is a great framework for our program: simple interactions between players can be analyzed analytically in both the temporal and spatial setting, thanks to recent advances by Durrett. 
%\item Second is role of OCY
%\item Third, it is a very elegant approach and comes with minimal complexity, in the spirit of Occam's razor. While initial models were simplistic, a lot of detail was added, but now reversal to a more global description.
%\item Primer to EGT
%\end{itemize}

\clearpage

\appendix

\section{Analysis of the Non-Spatial Game}\label{app_mean}
To analyze the replicator equation (\ref{e1}) on the simplex defined by $x(t)\geq 0$ and $x_1(t)+x_2(t)+x_3(t)=1$, we first analyze the three subgames that take place on the edges of the simplex: resorption-formation, resorption-quiescence, and formation-quiescence.  Denoting by
 $G=(g_{ij})$, $i,j\in \set{a,b}$ the pay-off matrix of a generic two-player game with players $a$ and $b$ of densities $x_a$ and $x_b$, respectively, the interior fixed point, if it exists, is located at $$\bar x_a=\frac{g_{12}-g_{22}}{g_{12}-g_{22}+g_{21}-g_{11}}, \quad \bar x_b=1-\bar x_a.$$

{\it Subgame 1: Resorption (r) vs Formation (f).} From \eqref{Gmatrix1}, the subgame pay-off matrix between resorption and formation is 
\begin{align}
\left( \begin{array}{cc}
0 & \alpha_3  \\
\beta_3 & 0
 \end{array} \right),
 \end{align} 
with potential interior fixed point $ \bar x_{r}=\frac{\alpha_3}{\alpha_3+\beta_3}.$ Since there are no a priori restrictions on $\alpha_3$ and $\beta_3$ (see Table 1), we distinguish between four different cases. (i) $\alpha_3>0$ and $\beta_3>0$: in this case, there is an interior fixed point, and by noticing that $x_r$ evolves according to  \eqref{e1},
\begin{align}\label{evo}
\frac{dx_r}{dt}= x_r(1-x_r) \ob{\alpha_3- \ob{\alpha_3+\beta_3}  x_r},
\end{align}
we find that $\bar x_r$ is attracting. (ii)  $\alpha_3<0$ and $\beta_3<0$; in this case there is an interior fixed point, and by \eqref{evo} it is repulsive. (iii) $\alpha_3>0$ and $\beta_3<0$: in this case there is no interior fixed point, and resorption will take over. (iv) $\alpha_3<0$ and $\beta_3>0$: in this case there is no interior fixed point, and formation will take over.\\

{\it Subgame 2: Resorption (r) vs Quiescence (q).}  The pay-off matrix of this subgame is
\begin{align}
\left( \begin{array}{cc}
0 & \beta_2  \\
\alpha_2 & 0
 \end{array} \right).
 \end{align} Recalling that $\alpha_2<0$, $\beta_2>0$ (Table 1), we find $\bar x_{r}=\frac{\beta_2}{\beta_2+\alpha_2} \notin (0,1)$, which means there is no interior fixed point. In addition, the evolution equation for resorption is 
\begin{align}\label{evo2}
\frac{dx_r}{dt}= x_r(1-x_r) \ob{\beta_2- \ob{\beta_2+\alpha_2}  x_r},
\end{align}  
which means $\bar x_r=1$ is globally attracting. This is consistent with the biology because as long as there are active osteoclasts attached to the bone matrix, the latter should be completely resorbed.\\

{\it Subgame 3: Formation (f) vs Quiescence (q).} The pay-off matrix is 
\begin{align}
\left( \begin{array}{cc}
0 & \alpha_1  \\
\beta_1 & 0
 \end{array} \right), 
 \end{align} 
and the evolution equation for formation is
\begin{align}\label{evo2}
\frac{dx_f}{dt}= x_f(1-x_f) \ob{\alpha_1- \ob{\alpha_1+\beta_1}  x_f}.
\end{align} 
Due to the parameter restrictions  $\alpha_1<0$ and $\beta_2>0$ from  \eqref{Gmatrix1}, there is no interior fixed point: $\bar x_{f}=\frac{\alpha_1}{\alpha_1+\beta_2} \notin (0,1)$. We note that the resulting attractive fixed point at $\bar x_{f}=0$ is consistent with the biology which requires zones of formation to produce new bone. \\   
 
Now that we have a complete understanding of the subgame dynamics, we can investigate the three-player game. An important concept in this analysis is the notion of {\it invadability} of edge fixed points \cite{durrett2002mutual}, which ascertains whether a small addition of player 3 can invade the edge equilibrium between players 1 and 2 or not. Because Durrett \cite{durrett2014spatial} previously characterized the dynamic regimes relevant to this analysis, we follow his notation and refer to his proofs where possible.  On occasion, we will also refer to the work of Bomze \cite{bomze1983lotka,bomze1995lotka} who has provided a complete characterization of the the replicator dynamics on the simplex.\\

Due to the parameter restrictions listed in Table 1, there are a total of seven different dynamic regimes to be discussed below. The corresponding partition of parameter space in the plane spanned by $\alpha_3$ and $\beta_3$, together with examples of trajectories for all seven cases, are shown in Figure \ref{fig:Figure_3}.\\

\noindent {\bf Case 1.} $\alpha_3>0$ and $\beta_3>0$. On the resorption-formation edge of the simplex, there is an attracting edge equilibrium at $
(\bar x_1, \bar x_2, \bar x_3)=\ob{\frac{\alpha_3}{\alpha_3+\beta_3}, \frac{\beta_3}{\alpha_3+\beta_3}}$  and $\ip{F} = F_1=F_2 =  \frac{\alpha_3\beta_3}{\alpha_3+\beta_3}$. We distinguish two subcases depending on whether quiescence can invade this equilibrium or not. According to \eqref{e1}, invadability is possible if  the expansion rate of quiescence exceeds the average expansion rate in the system, $F_3>\ip{F}$, which is equivalent to
\begin{align}\label{bone_inv}
 \alpha_2\alpha_3+\beta_1\beta_3-\alpha_3\beta_3>0.
 \end{align}  
 \begin{itemize}
 \item {\bf Case 1A.} If condition (\ref{bone_inv}) is satisfied, quiescence can invade the resorption-formation equilibrium, and there is an attracting interior fixed point, see Example 7.3 in \cite{durrett2014spatial}. 
  \item  {\bf Case 1B.}  If condition (\ref{bone_inv}) is not satisfied,  quiescence cannot invade the resorption-formation equilibrium, and there is no interior fixed point; all trajectories converge onto the resorption-formation equilibrium, see example 7.3.D in \cite{durrett2014spatial}.
 \end{itemize}
{\bf Case 2.} $\alpha_3<0$ and $\beta_3<0$. There is a repelling equilibrium on the resorption-formation edge. First, we note that the numerator of $\rho_1$ in \eqref{fpoints} is positive. The numerator of $\rho_3$ is positive if and only if condition (\ref{bone_inv}) is satisfied, which is equivalent to
\begin{align}\label{cond3}
 \frac{\alpha_2}{\beta_3}>1-\frac{\beta_1}{\alpha_3}.
 \end{align} Similarly, the numerator of $\rho_2$ is positive if and only if $\beta_2\beta_3+\alpha_1\alpha_2-\alpha_2\beta_2>0$, which is equivalent to 
 \begin{align}\label{cond2}
 \frac{\beta_3}{\alpha_2}<1-\frac{\alpha_1}{\beta_2}.
 \end{align}
\begin{itemize}
\item {\bf Case 2A.} If condition (\ref{cond3}) is satisfied, $\alpha_2/\beta_3>1$.  This implies that  (\ref{cond2}) is satisfied, too, and hence there  is an interior fixed point. This is the time-reversed case 15 in  \cite{bomze1983lotka}, which means the interior fixed point is unstable and the vertex $(1,0,0)$ is globally attracting.
\item {\bf Case 2B.} If condition (\ref{cond3}) is not satisfied, the numerator of $\rho_3$ is negative, whereas the one of  $\rho_1$ remains positive. In consequence, there is no interior fixed point.  This corresponds to the time-reversal of case 41 in \cite{bomze1983lotka}, and $(1,0,0)$ is again the global sink.\end{itemize} 
{\bf Case 3.} $\alpha_3>0$ and $\beta_3<0$. In this case, there are no edge fixed points, and resorption will take over, see Example 7.4.A in \cite{durrett2014spatial}. \\

{\bf Case 4.} $\alpha_3<0$ and $\beta_3>0$. In this case, the game matrix is a generalized Rock-Paper-Scissors game. There is an interior fixed point, and its stability is determined by the sign of $\Delta =  \beta_1\beta_2 \beta_3+\alpha_1\alpha_2\alpha_3$, see example 7.4 in \cite{durrett2014spatial}.
\begin{itemize}
\item {\bf Case 4A.} If $\Delta>0$, then there is an interior fixed point, and the solutions spiral inwards toward the fixed point.
\item {\bf Case 4B.} If $\Delta<0$, then the system is unstable and the boundary of the simplex is a limit cycle.
\end{itemize}

Note that if $\Delta=0$, there is a 1-parameter family of periodic orbits, but this is on a set of measure zero in parameter space, so we are not concerned with this case. 

%%%%%%%%%%%%%%%%%%%%%%%%%%%
\section{Analysis of the Spatial Game}\label{app_spatial}
%%%%%%%%%%%%%%%%%%%%%%%%%%%

To characterize the dynamics of the spatial game, we need to analyze the limiting behavior of the system in the weak selection limit \eqref{birthdeath}. To this end, we start by classifying the dynamic regimes of the well-mixed replicator dynamics for the pay-off matrix $H$. In other words, we study the behavior of equation \eqref{e1} where we replace $G$ by $H$. Analyzing first the embedded two-player games as in the non-spatial scenario (see Appendix A), it is straightforward to establish that quiescence dominates formation, and resorption dominates quiescence. The outcome of the resorption-formation game depends on the respective signs of $h_{12}$ and $h_{21}$, which are determined by
$$ h_{12}>0 \iff \alpha_3>\beta_3 \frac{\theta}{1+\theta} $$ and
$$ h_{21}>0 \iff \alpha_3<\beta_3 \frac{1+\theta}{\theta}. $$ 

\noindent {\bf Case 1.} If $h_{12}, h_{21}>0$, then as in Case 1 of the non-spatial game, there is an attracting fixed point on the resorption-formation edge, and we are interested in the invadability of this edge equilibrium. The latter is determined by the invadability condition
\begin{align}\label{cond123}
h_{12}h_{31}+h_{21}h_{32}>h_{12}h_{21}.
\end{align}
\begin{itemize}
\item {\bf Case 1A.}  If condition (\ref{cond123}) is satisfied, the resorption-formation equilibrium of the H-matrix replicator dynamics is invadable, and there is an interior attracting fixed point for the spatial game, see Example 7.3 in Section 8.3 of \cite{durrett2014spatial}.
%We note here that, because $h_{12}h_{21}>0$, a necessary condition for coexistence is $$\frac{h_{21}}{h_{12}}> \frac{|h_{31}|}{h_{32}}.$$
\item {\bf Case 1B.} If condition (\ref{cond123}) is not satisfied, the resorption-formation equilibrium is not invadable, and the H-matrix replicator dynamics do not admit coexistence of all three species. We conjecture that the same conclusion holds for the spatial game, but we are not able to explicitly prove this assertion. Nevertheless, simulations corroborate the conjecture, see Figure \ref{fig:Figure_3}B.
\end{itemize}
\noindent {\bf Case 2.} If $h_{12},h_{21}<0$, then the resorption-formation subgame has a repelling fixed point, with two possible cases:
\begin{itemize}
\item {\bf Case 2A.} If (\ref{cond123}) is satisfied, the boundary equilibrium is invadable, and there is no interior fixed point for the H-matrix replicator dynamics. In the spatial game, we conjecture takeover off resorption, and simulations support this hypothesis, see Figure \ref{fig:Figure_3}B.
\item {\bf Case 2B.} If  (\ref{cond123}) is not satisfied, the boundary equilibrium is not invadable, and there is an unstable interior fixed point. Again, we conjecture takeover by resorption in the spatial game, and corroborate the conjecture by simulation, see Figure \ref{fig:Figure_3}B.
\end{itemize}

\noindent {\bf Case 3.} If $h_{12}>0$ and $h_{21}<0$, then the resorption-formation subgame is dominated by resorption. We conjecture take-over by resorption in the spatial game and corroborate this by simulation, see Figure \ref{fig:Figure_3}B.\\

\noindent {\bf Case 4.} If $h_{12}<0$ and $h_{21}>0$ then the H-matrix dynamics constitute a generalized Rock-Paper-Scissor game, and the outcome of the temporal dynamics is determined by the sign of $\Delta:= h_{13}h_{21}h_{32}+h_{12}h_{23}h_{31}.$
\begin{itemize}
\item {\bf Case 4A.} If $\Delta>0$, there is an attracting interior fixed point for the H-matrix replicator dynamics. While it seems intuitive that this leads to coexistence in the spatial case a proof is out of reach. We corroborated our hypothesis by simulation, see Figures \ref{fig:Figure_3}B and \ref{fig:Figure_3_new}.
\item {\bf Case 4B.} If  $\Delta<0$, then the interior fixed point is unstable, and the solutions to the H-matrix replicator dynamics spiral outwards (with the boundary of the simplex as a limit-cycle). Following the discussion of non-hierarchical competition models by Durrett and Levin  \cite{durrett1998spatial}, we conjecture long-time coexistence in the spatial game. Simulations corroborated this hypothesis, see Figures \ref{fig:Figure_3}B and \ref{fig:Figure_3_new}.
\end{itemize}

\clearpage

\bibliographystyle{vancouver}
	\bibliography{bonegame}

\begin{thebibliography}{10}

\bibitem{robling2006biomechanical}
Robling AG, Castillo AB, Turner CH.
\newblock Biomechanical and molecular regulation of bone remodeling.
\newblock Annu Rev Biomed Eng. 2006;8:455--498.

\bibitem{raggatt2010cellular}
Raggatt LJ, Partridge NC.
\newblock Cellular and molecular mechanisms of bone remodeling.
\newblock J Biol Chem. 2010;285(33):25103--25108.

\bibitem{feng2011disorders}
Feng X, McDonald JM.
\newblock Disorders of bone remodeling.
\newblock Annu Rev Pathol. 2011;6:121.

\bibitem{lemaire2004modeling}
Lemaire V, Tobin FL, Greller LD, Cho CR, Suva LJ.
\newblock Modeling the interactions between osteoblast and osteoclast
  activities in bone remodeling.
\newblock J Theor Biol. 2004;229(3):293--309.

\bibitem{komarova2003mathematical}
Komarova SV, Smith RJ, Dixon SJ, Sims SM, Wahl LM.
\newblock Mathematical model predicts a critical role for osteoclast autocrine
  regulation in the control of bone remodeling.
\newblock Bone. 2003;33(2):206--215.

\bibitem{komarova2005mathematical}
Komarova SV.
\newblock Mathematical model of paracrine interactions between osteoclasts and
  osteoblasts predicts anabolic action of parathyroid hormone on bone.
\newblock Endocrinology. 2005;146(8):3589--3595.

\bibitem{pivonka2008model}
Pivonka P, Zimak J, Smith DW, Gardiner BS, Dunstan CR, Sims NA, et~al.
\newblock Model structure and control of bone remodeling: A theoretical study.
\newblock Bone. 2008;43(2):249--263.

\bibitem{buenzli2012modelling}
Buenzli PR, Pivonka P, Gardiner BS, Smith DW.
\newblock Modelling the anabolic response of bone using a cell population
  model.
\newblock Journal of theoretical biology. 2012;307:42--52.

\bibitem{ji2012novel}
Ji B, Genever PG, Patton RJ, Putra D, Fagan MJ.
\newblock A novel mathematical model of bone remodelling cycles for trabecular
  bone at the cellular level.
\newblock Biomech Model Mechan. 2012;p. 1--10.

\bibitem{pivonka2010mathematical}
Pivonka P, Komarova SV.
\newblock Mathematical modeling in bone biology: From intracellular signaling
  to tissue mechanics.
\newblock Bone. 2010;47(2):181--189.

\bibitem{webster2011silico}
Webster D, M{\"u}ller R.
\newblock In silico models of bone remodeling from macro to nano -- from organ
  to cell.
\newblock Wiley Interdiscip Rev Syst Biol Med. 2011;3(2):241--251.

\bibitem{parfitt1994osteonal}
Parfitt A.
\newblock Osteonal and hemi-osteonal remodeling: The spatial and temporal
  framework for signal traffic in adult human bone.
\newblock J Cell Biochem. 1994;55(3):273--286.

\bibitem{khosla2001minireview}
Khosla S.
\newblock Minireview: The OPG/RANKL/RANK system.
\newblock Endocrinology. 2001;142(12):5050--5055.

\bibitem{ryser2008mathematical}
Ryser MD, Nigam N, Komarova SV.
\newblock Mathematical Modeling of Spatio-Temporal Dynamics of a Single Bone
  Multicellular Unit.
\newblock J Bone Miner Res. 2008;24(5):860--870.

\bibitem{ryser2010cellular}
Ryser MD, Komarova SV, Nigam N.
\newblock {The Cellular Dynamics of Bone Remodeling: A Mathematical Model}.
\newblock SIAM J Appl Math. 2010;70:1899--1921.

\bibitem{ryser2012osteoprotegerin}
Ryser MD, Qu Y, Komarova SV.
\newblock Osteoprotegerin in Bone Metastases: Mathematical Solution to the
  Puzzle.
\newblock PLOS Comp Biol. 2012;8(10):e1002703.

\bibitem{buenzli2011spatio}
Buenzli PR, Pivonka P, Smith DW.
\newblock Spatio-temporal structure of cell distribution in cortical Bone
  Multicellular Units: A mathematical model.
\newblock Bone. 2011;48(4):918--926.

\bibitem{graham2012towards}
Graham JM, Ayati BP, Ramakrishnan PS, Martin JA.
\newblock Towards a new spatial representation of bone remodeling.
\newblock Math Biosci Eng. 2012;9(2):281.

\bibitem{buenzli2012investigation}
Buenzli PR, Jeon J, Pivonka P, Smith DW, Cummings PT.
\newblock Investigation of bone resorption within a cortical basic
  multicellular unit using a lattice-based computational model.
\newblock Bone. 2012;50(1):378--389.

\bibitem{van2008unified}
van Oers RFM, Ruimerman R, Tanck E, Hilbers PAJ, Huiskes R.
\newblock A unified theory for osteonal and hemi-osteonal remodeling.
\newblock Bone. 2008;42(2):250--259.

\bibitem{van2011simulations}
van Oers RFM, van Rietbergen B, Ito K, Huiskes R, Hilbers PAJ.
\newblock Simulations of trabecular remodeling and fatigue: Is remodeling
  helpful or harmful?
\newblock Bone. 2011;48(5):1210--1215.

\bibitem{badilatti2016large}
Badilatti SD, Christen P, Levchuk A, Marangalou JH, van Rietbergen B, Parkinson
  I, et~al.
\newblock Large-scale microstructural simulation of load-adaptive bone
  remodeling in whole human vertebrae.
\newblock Biomech Model Mech. 2016;15(1):83--95.

\bibitem{cox2011analysis}
Cox L, Van~Rietbergen B, Van~Donkelaar C, Ito K.
\newblock Analysis of bone architecture sensitivity for changes in mechanical
  loading, cellular activity, mechanotransduction, and tissue properties.
\newblock Biomech Model Mech. 2011;10(5):701--712.

\bibitem{scheiner2013coupling}
Scheiner S, Pivonka P, Hellmich C.
\newblock Coupling systems biology with multiscale mechanics, for computer
  simulations of bone remodeling.
\newblock Comput Method Appl M. 2013;254:181--196.

\bibitem{nakashima2011evidence}
Nakashima T, Hayashi M, Fukunaga T, Kurata K, Oh-hora M, Feng JQ, et~al.
\newblock Evidence for osteocyte regulation of bone homeostasis through RANKL
  expression.
\newblock Nat Med. 2011;.

\bibitem{xiong2011matrix}
Xiong J, Onal M, Jilka RL, Weinstein RS, Manolagas SC, O'Brien CA.
\newblock Matrix-embedded cells control osteoclast formation.
\newblock Nat Med. 2011;.

\bibitem{smith1982evolution}
Smith JM.
\newblock Evolution and the Theory of Games.
\newblock Cambridge University Press; 1982.

\bibitem{frey2010evolutionary}
Frey E.
\newblock Evolutionary game theory: Theoretical concepts and applications to
  microbial communities.
\newblock Physica A. 2010;389(20):4265--4298.

\bibitem{hammerstein1994game}
Hammerstein P, Selten R, et~al.
\newblock Game theory and evolutionary biology.
\newblock Handbook of game theory with economic applications. 1994;2:929--993.

\bibitem{nowak2006evolutionary}
Nowak MA.
\newblock Evolutionary Dynamics.
\newblock Harvard University Press; 2006.

\bibitem{broom2013game}
Broom M, Rychtar J.
\newblock Game-Theoretical Models in Biology.
\newblock Taylor \& Francis; 2013.

\bibitem{cox2013voter}
Cox JT, Durrett R, Perkins EA.
\newblock Voter Model Perturbations and Reaction Diffusion Equations.
\newblock Amer Mathematical Society; 2013.

\bibitem{durrett2014spatial}
Durrett R.
\newblock Spatial evolutionary games with small selection coefficients.
\newblock Electron J Probab. 2014;19.

\bibitem{eriksen1992cellular}
Eriksen E, Kassem M.
\newblock The cellular basis of bone remodeling.
\newblock Triangle. 1992;31(2):45--57.

\bibitem{liggett2013stochastic}
Liggett TM.
\newblock Stochastic interacting systems: contact, voter and exclusion
  processes. vol. 324.
\newblock Springer Science \& Business Media; 2013.

\bibitem{raisz1999physiology}
Raisz LG.
\newblock Physiology and pathophysiology of bone remodeling.
\newblock Clin Chem. 1999;45(8):1353--1358.

\bibitem{durrett1998spatial}
Durrett R, Levin S.
\newblock Spatial aspects of interspecific competition.
\newblock Theor Popul Biol. 1998;53(1):30--43.

\bibitem{durrett1994importance}
Durrett R, Levin S.
\newblock The importance of being discrete (and spatial).
\newblock Theor Popul Biol. 1994;46(3):363--394.

\bibitem{dingli2009cancer}
Dingli D, Chalub F, Santos F, Van~Segbroeck S, Pacheco J.
\newblock Cancer phenotype as the outcome of an evolutionary game between
  normal and malignant cells.
\newblock Brit J Cancer. 2009;101(7):1130--1136.

\bibitem{graham2013role}
Graham JM, Ayati BP, Holstein SA, Martin JA.
\newblock The role of osteocytes in targeted bone remodeling: a mathematical
  model.
\newblock PloS One. 2013;8(5):e63884.

\bibitem{buenzli2015osteocytes}
Buenzli PR.
\newblock Osteocytes as a record of bone formation dynamics: A mathematical
  model of osteocyte generation in bone matrix.
\newblock J Theor Biol. 2015;364:418--427.

\bibitem{buenzli2015quantifying}
Buenzli PR, Sims NA.
\newblock Quantifying the osteocyte network in the human skeleton.
\newblock Bone. 2015;75:144--150.

\bibitem{hellmich2004mineral}
Hellmich C, Barth{\'e}l{\'e}my JF, Dormieux L.
\newblock Mineral--collagen interactions in elasticity of bone
  ultrastructure--a continuum micromechanics approach.
\newblock Eur J Mech A - Solid. 2004;23(5):783--810.

\bibitem{hellmich2004can}
Hellmich C, Ulm FJ, Dormieux L.
\newblock Can the diverse elastic properties of trabecular and cortical bone be
  attributed to only a few tissue-independent phase properties and their
  interactions?
\newblock Biomech Model Mech. 2004;2(4):219--238.

\bibitem{durrett2002mutual}
Durrett R.
\newblock Mutual invadability implies coexistence in spatial models. vol. 740.
\newblock American Mathematical Soc.; 2002.

\bibitem{bomze1983lotka}
Bomze IM.
\newblock Lotka-Volterra equation and replicator dynamics: a two-dimensional
  classification.
\newblock Biol Cybern. 1983;48(3):201--211.

\bibitem{bomze1995lotka}
Bomze IM.
\newblock Lotka-Volterra equation and replicator dynamics: new issues in
  classification.
\newblock Biol Cybern. 1995;72(5):447--453.

\bibitem{zou2001tumor}
Zou W, Hakim I, Tschoep K, Endres S, Bar-Shavit Z.
\newblock Tumor necrosis factor-$\alpha$ mediates RANK ligand stimulation of
  osteoclast differentiation by an autocrine mechanism.
\newblock J Cell Biochem. 2001;83(1):70--83.

\bibitem{tani1999autocrine}
Tani-Ishii N, Tsunoda A, Teranaka T, Umemoto T.
\newblock Autocrine regulation of osteoclast formation and bone resorption by
  IL-1$\alpha$ and TNF$\alpha$.
\newblock J Dent Res. 1999;78(10):1617--1623.

\bibitem{quinn2001transforming}
Quinn JM, Itoh K, Udagawa N, H{\"a}usler K, Yasuda H, Shima N, et~al.
\newblock Transforming growth factor $\beta$ affects osteoclast differentiation
  via direct and indirect actions.
\newblock J Bone Miner Res. 2001;16(10):1787--1794.

\bibitem{teitelbaum2000bone}
Teitelbaum SL.
\newblock Bone resorption by osteoclasts.
\newblock Science. 2000;289(5484):1504--1508.

\bibitem{erlebacher1998osteoblastic}
Erlebacher A, Filvaroff EH, Ye JQ, Derynck R.
\newblock Osteoblastic responses to TGF-$\beta$ during bone remodeling.
\newblock Mol Biol Cell. 1998;9(7):1903--1918.

\bibitem{canalis1996insulin}
Canalis E, Agnusdei D.
\newblock Insulin-like growth factors and their role in osteoporosis.
\newblock Calcified Tissue Int. 1996;58(3):133--134.

\bibitem{li2005sclerostin}
Li X, Zhang Y, Kang H, Liu W, Liu P, Zhang J, et~al.
\newblock Sclerostin binds to LRP5/6 and antagonizes canonical Wnt signaling.
\newblock J Biol Chem. 2005;280(20):19883--19887.

\bibitem{winkler2003osteocyte}
Winkler DG, Sutherland MK, Geoghegan JC, Yu C, Hayes T, Skonier JE, et~al.
\newblock Osteocyte control of bone formation via sclerostin, a novel BMP
  antagonist.
\newblock EMBO J. 2003;22(23):6267--6276.

\bibitem{ducy2000osteoblast}
Ducy P, Schinke T, Karsenty G.
\newblock The osteoblast: a sophisticated fibroblast under central
  surveillance.
\newblock Science. 2000;289(5484):1501--1504.

\bibitem{bonewald2007osteocytes}
Bonewald LF.
\newblock Osteocytes as dynamic multifunctional cells.
\newblock Ann NY Acad Sci. 2007;1116(1):281--290.

\end{thebibliography}

\end{document}